\documentclass[10pt,conference]{IEEEtran}
\IEEEoverridecommandlockouts
\usepackage{cite}
\usepackage{amsmath,amssymb,amsfonts}
\usepackage{algorithmic}
\usepackage{subfigure}
\usepackage{graphicx}
\usepackage{textcomp}
\usepackage{xcolor}
\usepackage{tcolorbox}
\usepackage{float}
\usepackage{adjustbox}
\usepackage{etc}
\usepackage{tabularx} 
\usepackage{array}
\usepackage{caption} 
\usepackage{moreverb}
\usepackage{multirow}
\usepackage{caption}
\usepackage{url}
\usepackage{balance}

\definecolor{lightergray}{RGB}{220, 220, 220}

\newcolumntype{L}{>{\raggedright\arraybackslash}p{0.5\textwidth}} 

\def\BibTeX{{\rm B\kern-.05em{\sc i\kern-.025em b}\kern-.08em
    T\kern-.1667em\lower.7ex\hbox{E}\kern-.125emX}}
\begin{document}
\title{Improved IR-based Bug Localization with Intelligent Relevance Feedback}

\author{\IEEEauthorblockN{ Asif Mohammed Samir}
\IEEEauthorblockA{
Dalhousie University \\
asifsamir@dal.ca}
\and
\IEEEauthorblockN{ Mohammad Masudur Rahman}
\IEEEauthorblockA{
Dalhousie University \\
masud.rahman@dal.ca}
}
\maketitle

\begin{abstract}
Software bugs pose a significant challenge during development and maintenance, and practitioners spend nearly 50\% of their time dealing with bugs. Many existing techniques adopt Information Retrieval (IR) to localize a reported bug using textual and semantic relevance between bug reports and source code. However, they often struggle to bridge a critical gap between bug reports and code that requires in-depth contextual understanding, which goes beyond textual or semantic relevance. In this paper, we present a novel technique for bug localization --\textit{BRaIn}-- that addresses the contextual gaps by assessing the relevance between bug reports and code with Large Language Models (LLM). It then leverages the LLM’s feedback (a.k.a., Intelligent Relevance Feedback) to reformulate queries and re-rank source documents, improving bug localization. We evaluate BRaIn using a benchmark dataset --Bench4BL-- and three performance metrics and compare it against six baseline techniques from the literature. Our experimental results show that BRaIn outperforms baselines by 87.6\%, 89.5\%, and 48.8\% margins in MAP, MRR, and HIT@K, respectively. Additionally, it can localize $\approx$52\% of bugs that cannot be localized by the baseline techniques due to the poor quality of corresponding bug reports. By addressing the contextual gaps and introducing Intelligent Relevance Feedback, BRaIn advances not only theory but also improves the IR-based bug localization. 
\end{abstract}

\begin{IEEEkeywords}
Bug Localization, Intelligent Relevance Feedback, Query Reformulation,
 Information Retrieval, LLM
\end{IEEEkeywords}


\section{Introduction}

Software bugs can cause major financial losses and lead to data breaches, security vulnerabilities, and operational disruptions \cite{itcisq2020_stats, forbes2023_maintenance}. A recent software bug from Microsoft-owned CrowdStrike caused several hours of disruption in the U.S. airline industry, nearly halting operations and resulting in over \$10 billion in damages \cite{crowdstrike_microsoft, BI_crowdstrike}. Developers at the major IT companies, such as Microsoft and Google, have reported bug resolution as a top concern \cite{practitioners_bug_localization_study}. According to existing studies up to 50\% of the programming time is spent by developers on finding, understanding, and fixing software issues\cite{o2017debugging_stats, devops2024_dev_stats, britton2013reversible_stats}. Thus, any automated support to tackle these challenges can greatly benefit the developers.

Software bugs are submitted to bug-tracking systems (e.g., Bugzilla, JIRA) as bug reports, which might capture crucial hints for resolving software-related issues. Developers often rely on these reports to trace the origin of bugs in the code. 
However, the content and quality of bug reports can vary significantly based on their submitters' level of expertise and articulation skills. In particular, there might be variations in word choice and presence of technical terms\cite{furnas1987vocabulary, blizzard}.
Such variations pose challenges for developers when pinpointing the root cause of defects, even for seasoned practitioners\cite{rahman2021forgotten}. To address these challenges, there has been significant research targeting the detection or localization of software bugs over the last few decades.

Researchers have presented two major categories of methods to automatically localize software bugs: program spectrum analysis and Information Retrieval. First, spectrum-based methods rely on program execution traces for fault localization. However, the execution traces are not always readily accessible, which makes these methods less scalable \cite{spectra1, spectra2}. On the other hand, Information Retrieval (IR)-based methods use overlapping terms or keywords between bug reports and source code to localize bugs \cite{ir_bug_localization1, ir_bug_localization3_topic, ir_bug_localization2_spectra, ir_localization_ml_dl, ir_rl_reformulation_1}. They are lightweight and scalable. However, they also struggle with the \textit{vocabulary mismatch problems}\cite{furnas1987vocabulary} and may not always deliver satisfactory results due to sporadic term matching. Researchers have also incorporated historical data from past bug reports, code change history, past bug fixes, and bug recurrences\cite{ir_ml_amalgam, bl_code_change}. Although these enhancements have been reported to improve the performance of the IR-based methods in localizing bugs, a recent study\cite{bench4bl} suggests that they do not significantly outperform the previous methods.

Recent IR-based techniques focus on search queries and attempt to improve their queries by capturing syntactic, co-occurrence, and hierarchical dependencies among the words in bug reports \cite{blizzard, rahman2021forgotten, chaparro2017_ob_eb, sisman2012incorporating}. However, these methods only use terms found in bug reports, which could be poorly written or insufficient\cite{rahman2021forgotten}. As a result, they frequently fail to bridge the gap between natural language from bug reports and programming code from a project when searching for software bugs. To address this issue, several techniques attempt to enhance queries with relevant terms extracted from source documents through relevance feedback mechanisms\cite{PRF_generic_1, PRF_generic_2, PRF_generic_3, prf_kim_2, prf_concept_loc_3, prf_haiduc_4, blizzard}. However, the majority of these techniques naively consider the top few documents (based on textual similarity) as relevant, overlooking the need for a comprehensive understanding of the code. As a result, they may not always capture the most meaningful terms from source code for their search queries. \cite{prf_kim_2, chaparro2017_ob_eb}. Thus, the existing IR-based techniques for bug localization suffer from two major challenges as follows.


\textbf{(a) Relevance feedback against search queries might not be always relevant:}
Gay et al. \cite{prf_haiduc_4} proposed a manual, iterative approach that leverages relevance feedback from developers and constructs queries to search for buggy source documents. In contrast, Sisman et al. \cite{prf_sisman_kak_1} and Kim et al. \cite{prf_kim_2} select the top few documents as relevant (a.k.a., pseudo relevance feedback) and leverage the feedback to improve their search queries. However, these techniques rely heavily on textually similar documents, which may not be always relevant, especially when dealing with source code and bug reports. Thus, a deeper understanding of both bug reports and source code is warranted to improve the relevance feedback mechanism and the subsequent steps of Information Retrieval (e.g., query reformulation, retrieval).

\textbf{(b) Textual and semantic relevance might not be sufficient:}
Bug reports contain not only natural language texts but also technical jargons, commit diffs, stack traces, and program elements \cite{blizzard}. These artifacts describe the context and symptoms of encountered bugs \cite{chaparro2017_ob_eb}. Since natural language is loosely structured, it can introduce ambiguity by expressing the same idea in various ways \cite{furnas1987vocabulary}. Similarly, programming languages are more structured yet allow syntactically diverse expressions (e.g., iterative vs. functional approaches) and arbitrary naming conventions \cite{codeNaturalness, conciseConsistent, prf_sisman_kak_1}. This flexibility can result in textual mismatches, where keywords or key phrases in the bug report (e.g., ``\textit{download failed}") do not directly match the identifiers in the code (e.g., \texttt{fetchResource}). At the same time, semantic mismatches can arise when a problem encountered in the bug report does not correspond to the programming task implemented in the code. For example, the encountered problem -- ``\textit{download failed}" -- might not align well with the task -- ``\textit{HTTP/FTP operation task and get packets}" if the word-level semantics are considered only. It requires an understanding of the relationship between network operations and file downloading to establish their connection. In other words, to localize such bugs, automated tools or methods need to go beyond surface-level matching and comprehensively understand the context of an encountered problem as well as the functionality of the corresponding source code.

In this paper, we present a novel technique -- \textit{BRaIn} -- to support bug localization using Information Retrieval (IR) and Intelligent Relevance Feedback (IRF). Our approach overcomes the challenge of contextual understanding of software bugs using Transformer models\cite{attention_is_transformer} and localizes the bugs leveraging such understanding. First, BRaIn collects potentially buggy documents from a codebase by analyzing their contextual relevance to a bug leveraging transformer models (e.g., Mistral\cite{mistral}). That is, unlike the existing methods, our method captures more human-like feedback to a query (a.k.a., Intelligent Relevance Feedback). Second, it extracts appropriate terms from these documents and expands the original query by further leveraging the captured feedback. Finally, BRaIn reranks the source documents by executing the expanded query and employing the relevance feedback, providing a refined list of suspicious source documents.

We conducted experiments using 4,683 bug reports from a benchmark dataset-- Bench4BL\cite{bench4bl}. We evaluated the performance of our approach using three commonly used metrics: Mean Average Precision (MAP), Mean Reciprocal Rank (MRR), and HIT@K. Our approach is compared with six suitable baselines from the literature\cite{saha_bleuir, blizzard, prf_sisman_kak_1, dnnloc_ir, NextBug, RLocator}. BRaIn consistently outperformed existing techniques, showing 19.3\% and 87.6\% higher MAP scores than that of traditional and Machine Learning (ML)-based approaches, respectively. Similar gains were observed in MRR (17.5\% and 89.5\%) and HIT@10 (12.2\% and 48.8\%). These results underscore the effectiveness and superiority of our proposed technique in software bug localization.

Thus, this research make following contributions-
\begin{itemize}
\item A novel relevance feedback mechanism, namely \textit{Intelligent Relevance Feedback (IRF)}, that leverages the code understanding and reasoning capability of the LLM to offer useful feedback to a search query. It is neither naive like pseudo-relevance feedback nor costly like human feedback.
\item A novel approach -- \textit{BRaIn} -- that localizes software bugs using effective search queries and retrieval, supported by the Intelligent Relevance Feedback mechanism.
\item An extensive evaluation of \textit{BRaIn} using three commonly used metrics and $\approx$4.7K bug reports and comparison with six baselines from three areas of the literature.
\item A replication package\cite{replication_BRaIn} with a prototype, a curated dataset, and configuration details for third-party replication and reuse. 
\end{itemize}

\begin{table}[t]
        \caption{An Example of Bug Report and Search Techniques}
    \centering
    \begin{minipage}{0.95\linewidth} 
        \centering
        \begin{tabular}{p{0.19\linewidth} | p{0.63\linewidth} | p{0.04\linewidth}}
        \hline
        \multicolumn{2}{c}{Bug ID\# 2013 (Wildfly CORE)} & Rank \\ \hline \hline

        Title & Unable to access HTTP management interface secured by legacy LDAP realm. & 72 \\ \hline
        
        Description & When the HTTP management interface is secured with a legacy security realm using LDAP, the user is not prompted to provide credentials as should be in the case of BASIC HTTP authentication mechanism. Instead, a 403 HTTP status is returned directly. Users won't be able to migrate their current (6.4, 7.0) configuration to 7.1 without change. & 13 \\ \hline
        
         Baseline Query & Bug Title + Bug Description & 24 \\ \hline

        NextBug\cite{NextBug}  &  Cosine Similarity (Embedding + TF-IDF) between Bug Report and Source Documents  & 25 \\ \hline

        \textit{BRaIn} & Intelligent Relevance Feedback + Query Expansion + Scoring & 1 \\ \hline
    \end{tabular}
        \label{tab:bug_report} 
    \end{minipage}
    \\[0.3cm]
    \begin{minipage}{1\linewidth} 
        \centering
        \includegraphics[width=0.78\linewidth]{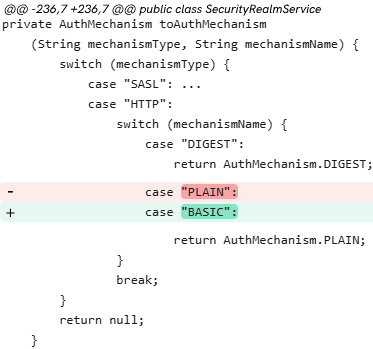}
        \captionof{figure}{Buggy Code with Diff} 
        \label{fig:buggy_code} 
    \end{minipage}
\end{table}

\section{Motivational Example}
\label{sect:motivational_ex}
In this section, we present a motivating example to demonstrate the benefits of our proposed technique for bug localization. Let us consider the example bug report in Table \ref{tab:bug_report} that discusses access problems to an LDAP server. The bug manifests as a failure in the authentication process, where the system returns an HTTP code of 403 (a.k.a., forbidden) instead of prompting for necessary credentials. This behavior results in a denial of access to the LDAP services and hinders any migration to a newer version of the services.


Fig. \ref{fig:buggy_code} presents the source code triggering the bug. The root cause of this bug is a subtle omission in the code handling authentication process. We see that the switch statement in the buggy version of code fails to account for the BASIC authentication type. Instead, it handles the PLAIN authentication type, which is semantically closer but not equivalent. On the other hand, the bug report mentions ``BASIC” HTTP authentication, which is not present in the target code. This terminology mismatch creates a disconnect between the high-level system behavior described in the bug report and the code level implementation, making the detection of bugs challenging. As a result, traditional text-based search methods perform poorly and retrieve the buggy code at 72\textsuperscript{nd}, 13\textsuperscript{th}, and 24\textsuperscript{th} positions when title, description, or their combination are used as queries, respectively (Table \ref{tab:bug_report}). Even after employing embedding-based semantic relevance, NextBug\cite{NextBug} struggles to link the code to the bug, placing it at 25\textsuperscript{th} position.

The above evidence suggests that an in-depth analysis involving contextual relevance is essential. A seasoned developer would recognize the missing clause of BASIC HTTP authentication in the code, although it is not explicitly stated in the bug report. By probing deeper, they would infer that the missing BASIC authentication is likely the root cause of the reported issue.

Large language models (e.g., Mistral) are exposed to a vast amount of data, including text and code. As a result, they can identify patterns (as humans do) and infer missing details, making them adept at handling tasks that require deep contextual understanding. As shown in Table \ref{tab:bug_report}, BRaIn leverages such capabilities to obtain intelligent feedback against its query, reformulates the query, and returns the buggy code as the topmost result by executing the query against an IR-based method (e.g., BM25\cite{robertson1995okapi_bm25}).

\begin{figure*}[ht]
    \centering
    \begin{tabular}{c} 
        \centering
        \includegraphics[width=0.99\linewidth]
        {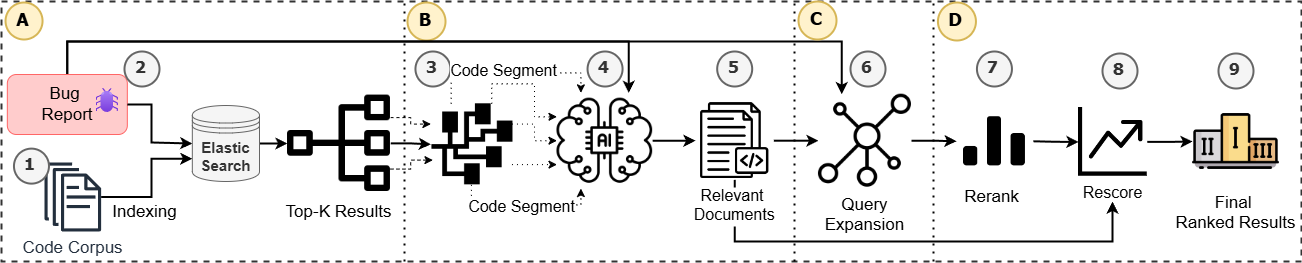}
    \end{tabular}

    \captionsetup{justification=centering}
    \caption{Schematic Diagram of \textit{BRaIn}:\\(A) Document Indexing \& Retrieval, (B) Intelligent Relevance Feedback, (C) Query Expansion, and (D) Bug Localization}
    \label{fig:schematic_diagram}
\end{figure*}

\section{Methodology}
Fig. \ref{fig:schematic_diagram} shows the schematic diagram of our proposed technique -- BRaIn -- for software bug localization. We discuss its different steps in the following section.

\subsection{Document Indexing and Retrieval}
\subsubsection{Indexing}
To detect software bugs using Information Retrieval (IR), the first step is to index the source code documents from a code repository. We chose Elasticsearch\cite{elasticsearch} for indexing due to its reliability, support for diverse data types, and easy integration with computing systems (e.g., cloud). We collected 45 subject systems from an existing benchmark dataset -- Bench4BL\cite{bench4bl} -- and indexed the source code (Step 1, Fig. \ref{fig:schematic_diagram}) from 684 buggy versions of these systems. Our idea was to detect a bug in the exact version of the software system stated in the corresponding bug report. During the indexing, we employed Elasticsearch’s default analyzer to perform common pre-processing operations (e.g., tokenization, lowercase conversion, and removal of stop words).

 \subsubsection{Retrieval of Potentially Buggy Documents using Textual Relevance}
To retrieve potentially buggy documents, we use bug reports (i.e., bug title and description) as queries (Step 2, Fig. \ref{fig:schematic_diagram}). When we pass these queries to Elasticsearch, it preprocesses them using the standard analyzer and returns the top-K (e.g., 50) results through query execution. To narrow down the search results, we also apply additional filters, such as system and version information from each bug report. Without these filters, the retrieved documents could be irrelevant or noisy. This step provides a set of source documents ranked by their textual relevance against a bug report by employing Elasticsearch’s default retrieval algorithm, Okapi BM25 \cite{robertson1995okapi_bm25}.

\subsection{Intelligent Relevance Feedback}
\label{sec:intelligent_relevance}
Once we have the results from Elasticsearch, we employ advanced prompt methods and Large Language Models (LLM) to determine the relevance between a bug report and each result (i.e., source code). LLMs have shown remarkable capabilities understanding natural language texts and source code\cite{llm_code_gen, llm_image, llm_speech}. We leverage their capabilities to capture intelligent relevance feedback against a query (a.k.a., bug report). To achieve this, we use prompt engineering, document segmentation, and finally relevance estimation as follows.

    \textbf{Prompt Engineering:}
\textit{Prompting} is a novel method that instructs the LLMs (e.g., Mistral) to generate meaningful responses without any expensive training \cite{prompting_general_1, prompting_general_2_design, prompting_general_3}. It involves crafting appropriate instructions to guide LLM outputs and make them applicable to different problem-solving tasks\cite{prompt_def_1, prompt_def_2, prompt_ex_t2image, prompt_ex_swe, prompt_ex_swe_2}. LLMs have been found to be effective with well-designed prompts that are clear, specific, and actionable \cite{prompting_general_2_design, prompt_survey}. Following the insights, we first developed a candidate prompt based on efficient prompt-building guidelines\cite{"prompting_how, prompting_how2}. Our goal was to determine whether a given code segment triggers a reported bug. It instructs the LLM to find the relevance, deliver the output in a JSON format, and act as a rational software engineer, incorporating the contextual information from the bug report and code segment.

To refine our candidate prompt, we employed SAMMO \cite{SAMMO}, a compile-time framework that optimizes prompts by exploring various configurations through mutation operations. We configured SAMMO with LLaMA-3\cite{llama} and used a small dataset of 20 bug reports with corresponding buggy code segments (ground truth) to guide our optimization process. SAMMO iteratively generated prompt variants by applying various modification operations to the candidate prompt with LLaMA’s assistance. In each iteration, we used LLaMA, bug reports and ground truth code to determine the fitness of each prompt and provide a performance update to SAMMO. Through an extensive search, SAMMO was able to find an optimized version of the prompt. Table \ref{tab:prompt-template} shows the optimized prompt template, used in the subsequent steps of our technique.

    \textbf{Segmentation:}
We divide the source code documents from Elasticsearch into smaller segments to determine their relevance using prompting and LLM (Step 3, Fig. \ref{fig:schematic_diagram}). According to an existing work\cite{segmentation_benefits}, breaking up texts into smaller segments helps the attention mechanism focus on specific parts, which could be useful for our relevance estimation task. In our work, we adopt a simple method to capture code segments rather than collecting program slices. The slicing methods often create slices that are either too small to capture meaningful context or too large, introducing irrelevant contexts and potentially exceeding the token limits of the LLMs\cite{code_slicing}. Therefore, we used a widely adopted library for static analysis -- JavaParser\cite{javaparser}, to extract code segments such as methods, constructors, interfaces, and enums from a document.

\textbf{Determining the relevance of code: }
To determine code relevance, we employ LLaMA\cite{llama}, Mistral\cite{mistral}, and Qwen\cite{qwen} models in a zero-shot setting, and provide a bug report and a code segment (e.g., method, constructor) as \textit{context} (Step 4, Fig. \ref{fig:schematic_diagram}), respecting the token limits of these models (e.g., 8,192). We used the optimized template in Table \ref{tab:prompt-template} for LLaMA and Qwen, consisting of three key elements: \textit{system, user, assistant}. On the other hand, the \textit{system} element does not apply to Mistral, and thus its prompt template was adapted accordingly. We use Hugging Face’s \cite{huggingface} AutoTokenizer to perform model-specific formatting of the prompt and vLLM \cite{vllm} to parallelize computations across the GPU in batches, increasing throughput. Each employed model against our context provides a response. For example, for the showcase bug report (Table \ref{tab:bug_report}) and corresponding buggy code (Fig. \ref{fig:buggy_code}), we obtained the JSON response \verb|{"relevance": "yes"}| from the Mistral. We also found a small number of cases where the outputs are malformed or incomplete JSON. In such instances, we perform string matching within the response (e.g., yes, no) to capture the relevance estimate of the code by the LLM. The relevance estimates of the code segments (collected from the results of Elasticsearch) serve as an intelligent feedback by the LLM to the original query. We coin this as \textit{Intelligent Relevance Feedback (IRF)}.

\begin{table}[h!]
\caption{Prompt Template for Relevance Feedback}
\centering
\begin{tabular}{|p{8cm}|}
\hline
\textbf{System:} \\
You are a helpful AI software engineer specializing in identifying buggy code segments given a bug report. Analyze the provided bug report and the JAVA code segment to determine if the code segment is responsible for causing the bug described in the bug report. You need to understand the functionality of the code segment and the details of the bug report to determine the relevance of the code segment to the bug report. \\

There are two possible outputs: `yes', `no'. \\
- `yes': The code is responsible for the bug described in the bug report. \\
- `no': The code is NOT responsible for the bug described in the bug report. \\

Provide your output in JSON format like this sample: \{``relevance'': ``yes''\}. \\

Act like a rational software engineer and provide output. Avoid emotion and extra text other than JSON. \\
\textbf{User:} \\
Analyze the following bug report and code segment: \\

Bug Report: \textless{}BUG REPORT\textgreater{} \\
Code Segment: \textless{}CODE SEGMENT\textgreater{} \\\\

Please determine if the code segment is responsible for the bug described in the bug report. \\

\textbf{Assistant:} \\
\hline
\end{tabular}
\label{tab:prompt-template}
\end{table}

\subsection{Query Expansion}
Using the Intelligent Relevance Feedback (IRF), we expand an original query (Step 6, Fig. \ref{fig:schematic_diagram}). Unlike the earlier work that relies on pseudo-relevance feedback\cite{prf_haiduc_4, prf_kim_2}, we choose the source code documents marked as relevant by the LLM for query expansion. In pseudo-relevance feedback, the top few documents retrieved by an IR method are naively considered as relevant and are used for query expansion. On the other hand, our underlying idea is that documents contextually relevant to the bug reports (i.e., IRF) provide terms that can complement the original query. We adapt an existing work of Rahman et al. \cite{coderank_rahman} to capture appropriate terms from the relevant source documents as follows.

First, we parse each of the source documents retrieved by Elasticsearch and extract class, method, and field signatures from them. These signatures capture the intent of the code, whereas the detailed implementation code could be noisy \cite{coderank_rahman}. We extract the signatures using a lightweight Python library -- Javalang\cite{javalang}. Next, we split camel case tokens from these signatures and turn them into textual phrases by combining their split tokens. We preprocess each of the phrases by filtering out stop words and programming keywords. Then, we construct a term graph $G(V, E)$ by encoding terms as vertices ($V$) and co-occurring terms as connecting edges ($E$). Subsequently, we apply the PageRank algorithm \cite{pagerank} in Eq. \ref{eq:pagerank} to the term graph to select the influential terms.

\begin{equation}
\label{eq:pagerank}
\small{
PR(V_i) = \frac{1 - d}{N} + d \sum_{V_j \in M(V_i)} \frac{PR(V_j)}{L(V_j)}
}
\end{equation}

Here, $PR(V_i)$ is the PageRank of vertex $V_i$. The term $d$ denotes the damping factor with a default set to 0.85. $N$ is the total number of vertices, while $M(V_i)$ is the set of vertices linked to $V_i$. Finally, $PR(V_j)$ and $L(V_j)$ denote the PageRank and outbound links of vertex $V_j$.

The algorithm assigns an initial score to each vertex ($V_i$) and iteratively updates it, prioritizing the vertices with higher connections. This process repeats until the scores stabilize or the algorithm reaches its maximum iteration limit (e.g., 100). Once the computation is done, we select the top-N (e.g., 10) weighted terms returned by the algorithm\cite{coderank_rahman}. Finally, we expand the original query (i.e., bug report) with these terms, complementing bug reports with contextually relevant terms from source code, leveraging intelligent relevance feedback.

\subsection{Bug Localization}
We leverage our expanded query above and Intelligent Relevance Feedback (IRF) from the LLM to localize the buggy source documents as follows.

\subsubsection{Reranking}
We determine the relevance between each result from Elasticsearch and our expanded query employing BM25 algorithm, and rerank them according to their relevance (Step 7, Fig. \ref{fig:schematic_diagram}). This step provides us with ranked results and their BM25 scores. The updated ranks could be useful since the expanded query contains more meaningful terms from the source documents.

\subsubsection{Rescoring}
We also enhance the ranking of source documents by incorporating IRF from LLM into their scores (Step 8, Fig. \ref{fig:schematic_diagram}). First, we normalize the BM25 scores of the retrieved documents using a softmax function \cite{franke2023_softmax}, producing a set of scores that add up to 1. The softmax function amplifies the differences among its input values exponentially, making their difference clearer. Given that the BM25 scores of the documents could have high variance, this allows the softmax function to highlight the textually relevant results. However, since textual relevance might not be sufficient, we also leverage the LLM's feedback against each result document. To incorporate this, we promote the relevant documents and penalize the irrelevant ones, marked by the LLM. This combined approach (Eq. \ref{eq:scoring function}) incorporates both textual and contextual relevance in the document ranking as follows.



\begin{equation}
\label{eq:scoring function}
\small
\textit{score}_i = \frac{e^{z_i}}{\sum_{j=1}^{n} e^{z_j}} \cdot r_i, \hspace{0.3cm} r_i = \begin{cases}
1, & \text{if } i^{\text{th}} \text{ doc is relevant} \\
0, & \text{otherwise}
\end{cases}
\end{equation}

Here, \(\textit{score}_i\) represents the score of the \(i^{\text{th}}\) document. The term \(r_i\) denotes the binary relevance feedback, indicating whether the document is relevant or not (details in Section \ref{sec:intelligent_relevance}). The terms \(z_i\) and \(z_j\) in the softmax function correspond to the BM25 scores of documents \(i\) and \(j\).

Finally, we rank the source documents based on their scores for their potential to be buggy (Step 9, Fig. \ref{fig:schematic_diagram}) and return the top-K (e.g., K=10) documents. Our scoring process aims to bridge the gap between bug reports and source code by incorporating deeper contextual understanding of the LLM beyond textual and semantic relevance (Table \ref{tab:bug_report}).

\begin{table}[ht]
    \caption{Dataset}
    \begin{minipage}[t]{0.28\textwidth} 
        \raggedright 
        \caption*{(a) Dataset Summary}
        \begin{tabular}{|c|c|c|}
            \hline
            Project & Systems & Bug Reports \\
            \hline
            \hline
            Spring  & 25 & 1,802 \\
            \hline
            Apache  & 25 & 1,802 \\
            \hline
            Wildfly & 5 & 806 \\
            \hline
            Commons & 8 & 507 \\
            \hline
            JBoss   & 1 & 9 \\
            \hline
            \hline
            Total & 42 & 4,683 \\
            \hline
        \end{tabular}
        \label{tab:dataset_summary}
    \end{minipage}%
    \begin{minipage}[t]{0.2\textwidth} 
        \centering
        \caption*{(b) Train-Test Split}
        \begin{tabular}{|c|c|c|}
            \hline
            Project & Train & Test \\
            \hline
            \hline
            Spring   & 1,429 & 373 \\
            \hline
            Apache   & 1,246 & 313 \\
            \hline
            Wildfly  & 643   & 163 \\
            \hline
            Commons  & 402   & 105 \\
            \hline
            JBoss    & 7     & 2   \\
            \hline
            \hline
            Total & 3,727 & 956 \\
            \hline
        \end{tabular}
        \label{tab:train-test_split}
    \end{minipage}
    \label{tab:dataset}
\end{table}


\section{Experiments}
We curate a dataset of $\approx$4.7K bug reports from the benchmark dataset Bench4BL and evaluate using three appropriate metrics from the relevant literature — Mean Average Precision (MAP), Mean Reciprocal Rank (MRR), and HIT@K (K=1, 5, 10)\cite{blizzard, saha_bleuir}. We experiment with three different LLMs and compare our solution --BRaIn-- against eight relevant baselines to place our work in the literature. Through our experiments, we answer three research questions as follows:

\begin{itemize}
\item \textbf{\textit{RQ$_1$}}: \textit{(a)} How does BRaIn perform in localizing software bugs? \textit{(b)} Does BRaIn enhance the localization of bugs that require changing multiple documents? \textit{(c)} Can it improve the localization of bugs that are reported poorly?
\item \textbf{\textit{RQ$_2$}}:  How do IRF-based query expansion and document ranking contribute to the performance of BRaIn?
\item \textbf{\textit{RQ$_3$}}: Can BRaIn outperform the relevant baseline techniques in bug localization?
\end{itemize}


\subsection{Dataset Construction}
\label{section:dataset_constraction}
In our experiment, we used the Bench4BL \cite{bench4bl}, a comprehensive benchmark dataset that contains 10,017 bug reports from 51 open-source systems, covering a total of 695 software versions.
Our initial assessment revealed that bug reports from the older systems lacked crucial versioning information, making them unsuitable for our study. 
Additionally, we could not accurately link some bug reports to their corresponding buggy code within the code repositories. Hence, we excluded these bug reports from our dataset.
We also found bug reports containing only stack traces without any textual descriptions of their bugs. 
We identified those bug reports using regular expressions\cite{blizzard} and excluded them from the dataset. Following these refinement steps, our final dataset comprised 4,683 bug reports from 42 different systems spanning across 684 versions. Table \ref{tab:dataset}-a summarizes our curated dataset.

To conduct our experiments and compare with deep learning-based baseline techniques, we used an optimal 80:20 dataset split\cite{dataset_split_ratio}. This split was done chronologically within each system to ensure that the training set consists of the older 80\% of the data, while the test set contains the newest 20\% to imitate a real-world scenario. Table \ref{tab:dataset}-b provides a summary of our training and test datasets.


\subsection{Evaluation Metrics}
\subsubsection*{\textbf{Mean Average Precision (MAP)}}

Precision@K indicates the precision for each instance of a buggy source document in the ranked list. Average Precision computes the average Precision@K for all buggy documents in relation to a specific search query. Consequently, Mean Average Precision (MAP) is obtained by averaging the Average Precision values across all queries (Q) within a dataset.

{\small
\vspace{-.5cm}
\begin{align*}
AP@K &= \frac{1}{|D|} \sum_{k=1}^{K} P_k \times B_k & \bigg| & \quad 
MAP &= \frac{1}{|Q|} \sum_{q=1}^{Q} AP@K_q
\end{align*}
}

Here, $AP@K$ computes average precision for top-$K$ results, where $P_k$ is precision at position $k$ and $B_k$ indicates if item $k$ is buggy (1) or not (0). $MAP$ averages this across all queries $q$ in dataset $Q$, with $D$ being the ground truth documents.

\subsubsection*{\textbf{Mean Reciprocal Rank (MRR)}}

Reciprocal Rank (RR) refers to the rank of the first relevant result retrieved by a technique. It is defined as the reciprocal of the rank of the first relevant source document within the ranked list for each query.

\vspace{-0.6cm}
{\small
\begin{align*}
RR_q &= \frac{1}{\textit{Rank of First Relevant Item}} & \bigg| & \quad 
MRR = \frac{1}{|Q|} \sum_{q=1}^{|Q|} RR_q
\end{align*}
}

Here, $MRR$ averages the Reciprocal Ranks ($RR_q$) across all queries $q$ in set $Q$.




\subsubsection*{\textbf{HIT@K}}

HIT@K \cite{saha_bleuir} measures the proportion of queries for which a technique retrieves at least one relevant document among the top-\( K \) results. Higher HIT@K values indicate better performance in bug localization techniques.

\vspace{-0.4cm}
{\small
\begin{align*}
HIT@K &= \frac{1}{|\mathcal{Q}|} \sum_{q=1}^{|Q|} \begin{cases} 
1, & r_q \in \mathcal{G} \\ 
0, & \text{otherwise} 
\end{cases} 
\end{align*}
}

Here, $r_q$ returns 1 if query $q$ has a ground truth item in the top-$K$ results (0 otherwise), where $Q$ is the set of all queries.

\subsection{Selection of LLM}
We select three Large Language Models (LLM) to design our techniques and conduct our experiments. We adopt three important criteria to select the models: (a) they should be open-source instruction models, (b) they need to have a quantized version to reduce computational demand, and (c) they should have similar numbers of parameters to allow for fair comparisons. Based on these criteria, we chose LLaMA-3 8B Instruct\cite{llama}, Mistral v0.3 7B Instruct\cite{mistral}, and Qwen 1.5 7B Chat\cite{qwen}. They were the top models from instruct category on the Huggingface Open LLM leaderboard\cite{llm_leaderboard} during July 2024. We use the 8 bit quantized GPTQ versions of these models\cite{GPTQ} to achieve computational efficiency and leverage vLLM\cite{vllm} for parallelization. We conducted the experiments on Nvidia V100 GPU-enabled machines with 16 GB vRAM in a cluster computing environment.

\begin{table}[!ht]
    \caption{Performance of BRaIn}
    \centering
    \begin{tabular}{|l|l|l|l|l|l|}
    \hline
        Techniques & MAP & MRR & HIT@1 & HIT@5 & HIT@10  \\ \hline
        \hline
        Elasticsearch & 0.484 & 0.513 & 0.413 & 0.647 & 0.732  \\ \hline \hline
        BRaIn (LLaMA) & 0.534 & 0.568 & 0.470 & 0.701 & 0.766  \\ \hline
        BRaIn (Mistral) & 0.537 & 0.571 & 0.469 & 0.709 & 0.781  \\ \hline
        BRaIn (Qwen) & 0.492 & 0.523 & 0.411 & 0.678 & 0.755  \\ \hline
    \end{tabular}
    \label{tab:rq1-a performance}
\end{table}

\begin{table*}[!ht]
    \caption{Performance of BRaIn against multi-document bugs}
    \centering
    \begin{tabular}{|c|c|c|c|c|c|c|c|c|c|c|c|c|c|}
    \hline
        \multirow{2}{*}{\shortstack{Changed \\ Documents}} & \multirow{2}{*}{\shortstack{Bug Report \\ Count}} & \multicolumn{3}{c|}{Elasticsearch (ES)} & \multicolumn{3}{c|}{LLaMA} & \multicolumn{3}{c|}{Mistral} & \multicolumn{3}{c|}{Qwen} \\ 
        & & MAP & MRR & HIT@10 & MAP & MRR & HIT@10 & MAP & MRR & HIT@10 & MAP & MRR & HIT@10 \\ \hline
        \hline  
        1 & 1,949 & 0.474 & 0.474 & 0.690 & 0.535 & 0.535 & 0.726 & 0.542 & 0.542 & 0.739 & 0.483 & 0.483 & 0.712 \\ \hline
        2 & 1,436 & 0.528 & 0.573 & 0.767 & 0.577 & 0.628 & 0.801 & 0.565 & 0.618 & 0.820 & 0.529 & 0.578 & 0.792 \\ \hline
        3 & 525 & 0.462 & 0.518 & 0.743 & 0.493 & 0.554 & 0.781 & 0.511 & 0.573 & 0.789 & 0.469 & 0.528 & 0.758 \\ \hline
        4$\geq$ & 773 & 0.445 & 0.501 & 0.765 & 0.480 & 0.551 & 0.793 & 0.492 & 0.554 & 0.811 & 0.461 & 0.520 & 0.794 \\ \hline
    \end{tabular}
    \label{tab:RQ1-b_multiple_changed}
\end{table*}

\subsection{Evaluating BRaIn}
\subsubsection*{\textbf{Answering RQ$_1$ - Performance of BRaIn}} 
We evaluate the performance of \textit{BRaIn} using Mean Average Precision (MAP), Mean Reciprocal Rank (MRR), and HIT@K against top 1, 5, and 10 results. Table \ref{tab:rq1-a performance} summarizes our performance details.

From Table \ref{tab:rq1-a performance}, we see that our proposed technique performs well in detecting the software bugs. BRaIn, powered by Mistral exhibits strong performance, with a Mean Average Precision (MAP) of 0.537. This indicates BRaIn's ability to rank the relevant documents (a.k.a., buggy source documents) higher than the irrelevant ones. Our technique achieves a Mean Reciprocal Rank (MRR) of 0.570 suggesting that the first relevant document is found within the top two positions. BRaIn (Mistral)’s HIT@1 score of 0.469 shows that, for nearly 47\% of bug reports, the most relevant document appears at the top position. BRaIn (Mistral) also performs well in HIT@5 and HIT@10, with approximately 71\% and 78\% of bug reports having at least one relevant buggy document found within the top 5 and top 10 positions, respectively. BRaIn (LLaMA) delivers nearly comparable results to BRaIn (Mistral), trailing by 1.9\% in HIT@10. Although BRaIn (Qwen) demonstrates decent performance, it lags behind both BRaIn (Mistral) and BRaIn (LLaMA) in all metrics. It achieves MAP and MRR scores of 0.492 and 0.523, which are about 9.1\% lower than BRaIn (Mistral)'s best performance in each metric.

According to our investigation, some bugs trigger changes to a single document during bug resolution, whereas others trigger changes to multiple documents.
We thus evaluate BRaIn's performance in localizing bugs that warrant changes across multiple source documents. Table \ref{tab:RQ1-b_multiple_changed} shows the performance of BRaIn, powered by three different LLMs, in terms of MAP, MRR, and HIT@10. We grouped bugs from our dataset into four categories based on the number of their changed documents: 1, 2, 3, and 4 or more. Our findings show that BRaIn performs strongest when paired with Mistral. For the 1,949 bugs requiring changes to a single document, BRaIn (Mistral) achieves a MAP of 0.542, representing a significant 14.3\% improvement over the baseline counterpart. The improvements extend to other metrics, with a 14.3\% increase in MRR and 7.1\% in HIT@10. BRaIn (Mistral) also excels in resolving bugs that require multiple document changes, achieving improvements of 7.0-10.6\% in MAP, 7.9-10.6\% in MRR, and 6.0-6.9\% in HIT@10. The other variants of BRaIn also outperform the Elasticsearch baseline in localizing bugs that require multiple document changes, achieving improvements of up to 9.2\% in MAP, 10.0\% in MRR, and 5.1\% in HIT@10.

\begin{figure}[t]
    \centering
    \begin{tabular}{cc} 
       \includegraphics[width=0.44\linewidth]{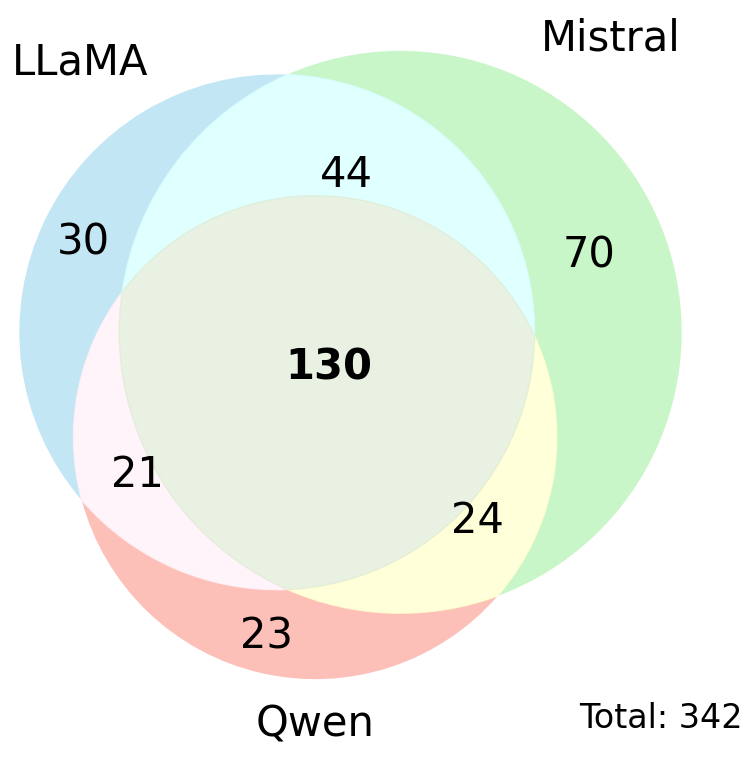}
        &
        \includegraphics[width=0.47\linewidth]{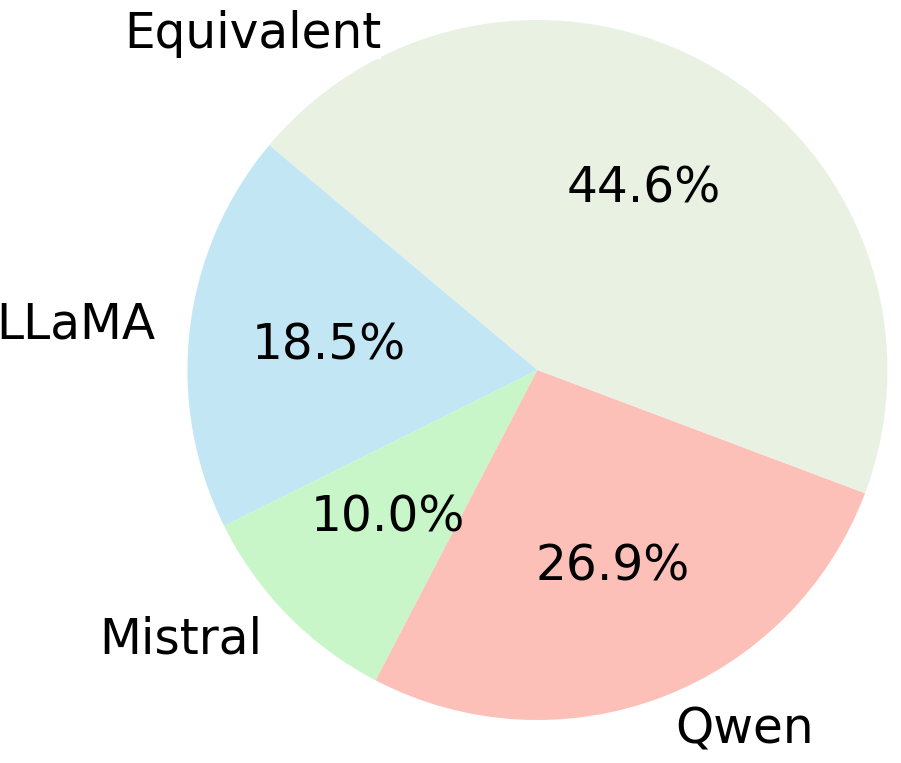}
        \\
        (a)
        &
        (b)
    \end{tabular}
    \caption{Performance of BRaIn with Low Quality Bug Reports}
    \label{fig.low quality}
\end{figure}

\label{section:rq1}
We also investigate how BRaIn performs in localizing bugs where the bug reports could be of low quality (Fig. \ref{fig.low quality}-a). According to existing literature \cite{rahman2021forgotten}, low-quality bug reports lack sufficient information and provide queries that cannot retrieve at least one relevant result within their top 10 positions. In our dataset, we identified 1,101 bug reports that fall into this category. Of these, 581 bug reports (i.e., queries) do not contain any ground truth within their top 50 results returned by Elasticsearch. These bug reports were not considered, which leaves us with 520 low-quality bug reports for our analysis. Our findings demonstrate BRaIn's promising results even with the low-quality reports. BRaIn (Mistral) emerges as the top performer, successfully localizing 268 bug reports (51.5\% of low-quality bug reports) within the top 10 results. BRaIn (LLaMA) and BRaIn (Qwen) follow, identifying 225 and 198 reports, respectively. Notably, all three models identified 130 bugs, with BRaIn (Mistral) uniquely localizing an additional 70 bugs, followed by BRaIn (LLaMA) and BRaIn (Mistral) with 30 and 23 bugs.
We also assess BRaIn's capability to detect the first relevant document (a.k.a., buggy source document), where 130 bug reports from above were considered (Fig. \ref{fig.low quality}-b). Interestingly, BRaIn (Qwen) outperformed the other variants in this metric, localizing 26.9\% of the bugs at the top positions, followed by BRaIn (LLAMA) at 18.5\% and BRaIn (Mistral) at 10\%.
All the findings above suggest BRaIn's ability to analyze, enhance, and localize bug reports, even with low-quality reports.


\begin{tcolorbox}[colback=lightergray,colframe=black,arc=1mm,boxrule=0.5pt]
\textbf{RQ1 Summary:} 
BRaIn significantly improves bug localization, particularly with Mistral, reaching a high MAP score of 0.537. This performance is due to BRaIn's effective handling of bug reports with up to $\approx$11\% multiple changed documents. Moreover, BRaIn demonstrates an impressive performance with limited information, successfully localizing $\approx$52\% of low-quality bug reports within the top 10 results where the baseline failed.
\end{tcolorbox}


\begin{table}[!ht]
    \caption{Impact of Query Expansion and Scoring}
    \centering
    \begin{tabular}{|c|c|c|c|c|c|c|}
    \hline
        \multirow{2}{*}{\shortstack{BRaIn \\ Components}} & \multicolumn{2}{c|}{BRaIn (LLaMA)} & \multicolumn{2}{c|}{BRaIn (Mistral)} & \multicolumn{2}{c|}{BRaIn (Qwen)} \\ 
        & MAP & MRR & MAP & MRR & MAP & MRR \\ \hline \hline
        \shortstack{Expansion + \\ Reranking} & 0.534 & 0.568 & 0.537 & 0.571 & 0.492 & 0.523 \\ \hline \hline
        \shortstack{Expansion + \\ No Reranking} & 0.498 & 0.568 & 0.498 & 0.569 & 0.496 & 0.567 \\ \hline
        \shortstack{No Expansion + \\ Reranking} & 0.518 & 0.574 & 0.520 & 0.573 & 0.489 & 0.529 \\ \hline
        
    \end{tabular}
    \label{tab:RQ2-a_expansionScoring}
\end{table}

\begin{table*}[!ht]
    \caption{Comparison Between BRaIn and Baseline Techniques}
    \begin{minipage}{0.58\textwidth}
        \centering
        \small 
        \caption*{ (a) Comparison with Non-ML Baselines}
        \begin{tabular}{|c|c|c|c||c|c|c|}
        \hline
            \multirow{2}{*}{\shortstack{Metrics}} & \multicolumn{3}{c||}{Traditional IR} & \multicolumn{3}{c|}{Relevance Feedback} \\ 
            & ES & BLUiR & Blizzard & Rocchio & Sysman-SCP & BRaIn \\ \hline \hline
            MAP & 0.484 & 0.450 & 0.506 & 0.489 & 0.472 & 0.537 \\ \hline
            MRR & 0.513 & 0.471 & 0.536 & 0.558 & 0.541 & 0.571 \\ \hline
            HIT@10 & 0.732 & 0.696 & 0.758 & 0.765 & 0.753 & 0.781 \\ \hline
        \end{tabular}
        \label{tab:comparison_with_BRaIn}
    \end{minipage}
    \hspace{0.00125\textwidth} 
    \begin{minipage}{0.38\textwidth}
        \small 
        \caption*{(b) Comparison with ML Baselines}
        \begin{tabular}{|c|c|c|c|c|}
        \hline 
            \multirow{2}{*}{\shortstack{Metrics}} & \multicolumn{4}{c|}{Machine Learning} \\ 
            & DNNLOC & RLocator & NextBug & BRaIn  \\ \hline \hline
            MAP & 0.283 & 0.488 & 0.469 & 0.531 \\ \hline
            MRR & 0.296 & 0.561 & 0.540 & 0.564 \\ \hline
            HIT@10 & 0.518 & 0.735 & 0.743 & 0.771 \\ \hline
        \end{tabular}
        \label{tab:ml_model_comparison}
    \end{minipage}
    
\end{table*}

\subsubsection*{\textbf{Answering RQ$_2$ - Contribution of Query Expansion and Document reranking}} BRaIn leverages Intelligent Relevant Feedback (IRF) to expand its original query and rerank the documents. Query expansion can improve bug localization by adding relevant keywords to an original query. Similarly, incorporating contextual understanding into document scoring can help go beyond just textual and semantic matching during document ranking. We examine the contribution of these two components (Table \ref{tab:RQ2-a_expansionScoring}) to BRaIn's performance as follows.

To determine the impact of query expansion in isolation, we evaluated BRaIn's performance without the reranking component (Table \ref{tab:RQ2-a_expansionScoring}). Interestingly, all BRaIn variants achieved similar MAP scores of around 0.49, falling short of optimal performance. For example, BRaIn (LLaMA) and BRaIn (Mistral) saw MAP scores decrease by 7.2\% and 7.8\%, respectively, while their MRR scores remained relatively stable. BRaIn (Qwen), on the other hand, showed a 7.8\% increase in MRR alongside a slight improvement in MAP. However, all BRaIn variants outperformed the Elasticsearch baseline by 10.5-10.9\% in MRR and 2.4-2.9\% in MAP.

In contrast, when reranking was applied isolately, MAP scores for BRaIn (LLaMA), BRaIn (Mistral), and BRaIn (Qwen) dropped by 3.0\%, 3.2\%, and 6.1\%, respectively, while their MRR scores remained consistent. Despite these decreases, all variants showed improvements of 3.1-11\% in MRR and 1.0-7.1\% in MAP over the Elasticsearch baseline.

These findings show how Intelligent Relevance Feedback improves query expansion and reranking. However, they also underscore the importance of the synergy between these components for optimal performance.

\begin{tcolorbox}[colback=lightergray,colframe=black,arc=1mm,boxrule=0.5pt]
\textbf{RQ2 Summary:} Query expansion and reranking individually decrease MAP and MRR by 3.0-7.8\%, yet both improve bug localization performance over the baseline by 2.4-11\% in these metrics. Their individual results highlight the contribution of Intelligent Relevance Feedback and underscore the importance of a synergistic combination in BRaIn.
\end{tcolorbox}

\subsubsection*{\textbf{Answering RQ$_3$ - Comparison with Basline Techniques}} 
To place our work in the literature, we compare BRaIn with relevant baseline techniques in terms of their MAP, MRR, and HIT@10. Given our methodology, we choose two types of baseline techniques -- IR methods\cite{saha_bleuir, blizzard, prf_haiduc_4} and deep learning based methods\cite{dnnloc_ir, NextBug, RLocator}. For comparison with baselines, we use BRaIn (Mistral) in our experiments since it is the best-performing variant of BRaIn.

To replicate the traditional IR-based bug localization with Elasticsearch baseline, we index all source documents of a repository and use bug reports (title + description) as queries. These queries are executed with Elasticsearch \cite{elasticsearch}, which retrieves relevant documents using the BM25 algorithm \cite{robertson1995okapi_bm25} and Boolean queries, with default parameters for k and b.
Other traditional approaches from literature-- BLUiR \cite{saha_bleuir} and Blizzard \cite{blizzard}-- use structured information from bug reports and source code for bug localization. BLUiR calculates suspiciousness scores using class names, method names, variable names, comments, and bug report elements (title, description), combining multiple searches into an overall score. Blizzard categorizes bug reports into three types and constructs text graphs from these reports to generate queries and retrieve relevant buggy source documents. For both approaches, we employ Apache Lucene\cite{lucene} for retrieval. We replicated these methods by adapting them from Bench4BL repository\cite{bench4bl} and Blizzard’s replication package\cite{BLIZZARD_github} from the authors. We compare BRaIn with these established IR-based techniques 
to validate our technique and place it in the literature.

Relevance feedback-based techniques like Rocchio\cite{rocchio_cambridge} and the Spatial Code Proximity (SCP) model\cite{prf_haiduc_4} aim to enhance bug localization by refining queries based on the results of initial searches. Rocchio is a widely-used relevance feedback technique for information retrieval \cite{rocchio_cambridge}. We leverage relevance feedback to reformulate queries and Apache Lucene to execute the queries and retrieve the documents. Our reformulated queries were optimized using $\alpha$, $\beta$, and $\gamma$ parameters \cite{rocchio_cambridge}. Similarly, we implemented Sisman et al.'s SCP model \cite{prf_haiduc_4} to reformulate queries based on term proximity within source code. It prioritizes terms that frequently co-occur within the same method or class, using the best parameters $w$, $x$, and $y$ suggested by the authors. We compare these techniques to our approach to highlight the importance of contextual understanding during relevance feedback of search queries.




Since Machine Learning (ML) techniques can capture complex patterns in data using non-linear relationships, we compare BRaIn against three ML-based techniques-- DNNLOC \cite{dnnloc_ir}, NextBug \cite{NextBug}, and RLocator \cite{RLocator}. DNNLOC combines multiple features— rVSM score\cite{ir_localization_lda_buglocator} for bug report-source code similarity, class name similarity, collaborative filtering, and bug report recency and frequency—and uses a neural network to predict suspiciousness scores to rank documents. NextBug employs Word2Vec \cite{word2vec} embeddings to capture semantic relations between bug reports and source code and thus to localize the buggy documents. In our experiments, we substituted Word2Vec with CodeT5 embeddings \cite{wang2021codet5} to capture more nuanced text-level semantic associations, as opposed to token-level. RLocator is a recent deep-learning technique that employs a reinforcement learning model, framed as a Markov Decision Process, to optimize ranking of buggy documents. We replicated DNNLOC and NextBug by following the respective authors' approaches and replicated RLocator using the authors' provided replication package on Zenodo\cite{RLocator_replication}. To ensure consistency with the authors' specifications, we replicated the methods using cross-validation.

Table \ref{tab:comparison_with_BRaIn}-a summarizes our comparison details with the baseline techniques. Among traditional IR-based approaches, Blizzard achieves a MAP score of 0.506, while Elasticsearch (ES) and BLUiR scores are 0.484 and 0.450. BRaIn outperforms them with a MAP of 0.537, achieving a maximum improvement of 19.3\% over these techniques. Similarly, BRaIn achieves notable gains in MRR and HIT@10, with increases of up to 17.5\% and 12.2\%. Among the IR based approaches that leverage relevance feedback, Rocchio’s algorithm achieves a MAP of 0.489, slightly above the baseline, while Sysman-SCP falls short by 2.5\%. BRaIn again leads here with the improvements in MAP, MRR, and HIT@10 of 13.8\%, 5.5\%, and 3.7\%, respectively. These results underscore the advantages of Intelligent Relevance Feedback, which uses contextual understanding over traditional techniques based on textual relevance for bug localization.


As shown in \ref{tab:ml_model_comparison}-b, BRaIn also outperforms the machine learning techniques that require training. We evaluated both BRaIn and the baseline techniques on the test set only to ensure a fair comparison. It should be noted that old bug reports and their corresponding code were used for training and the recent bugs and their corresponding code were used for testing. 
DNNLOC performs significantly lower with a MAP of 0.283, 87.6\% lower than BRaIn’s optimal score of 0.531. In comparison, RLocator and NextBug achieve MAP scores of 0.488 and 0.469, with BRaIn outperforming them by 8.8\% and 13.2\%, respectively. Similar improvements are observed for other metrics, with BRaIn showing 4.4-89.5\% improvements in MRR and 3.7-48.8\% in HIT@10. Such performance underscores the superiority of BRaIn's performance with Intelligent Relevance Feedback (IRF) compared to baseline techniques.

\begin{figure}[ht!]
    \centering
    \begin{tabular}{cc} 
        
        \includegraphics[width=0.8\linewidth]{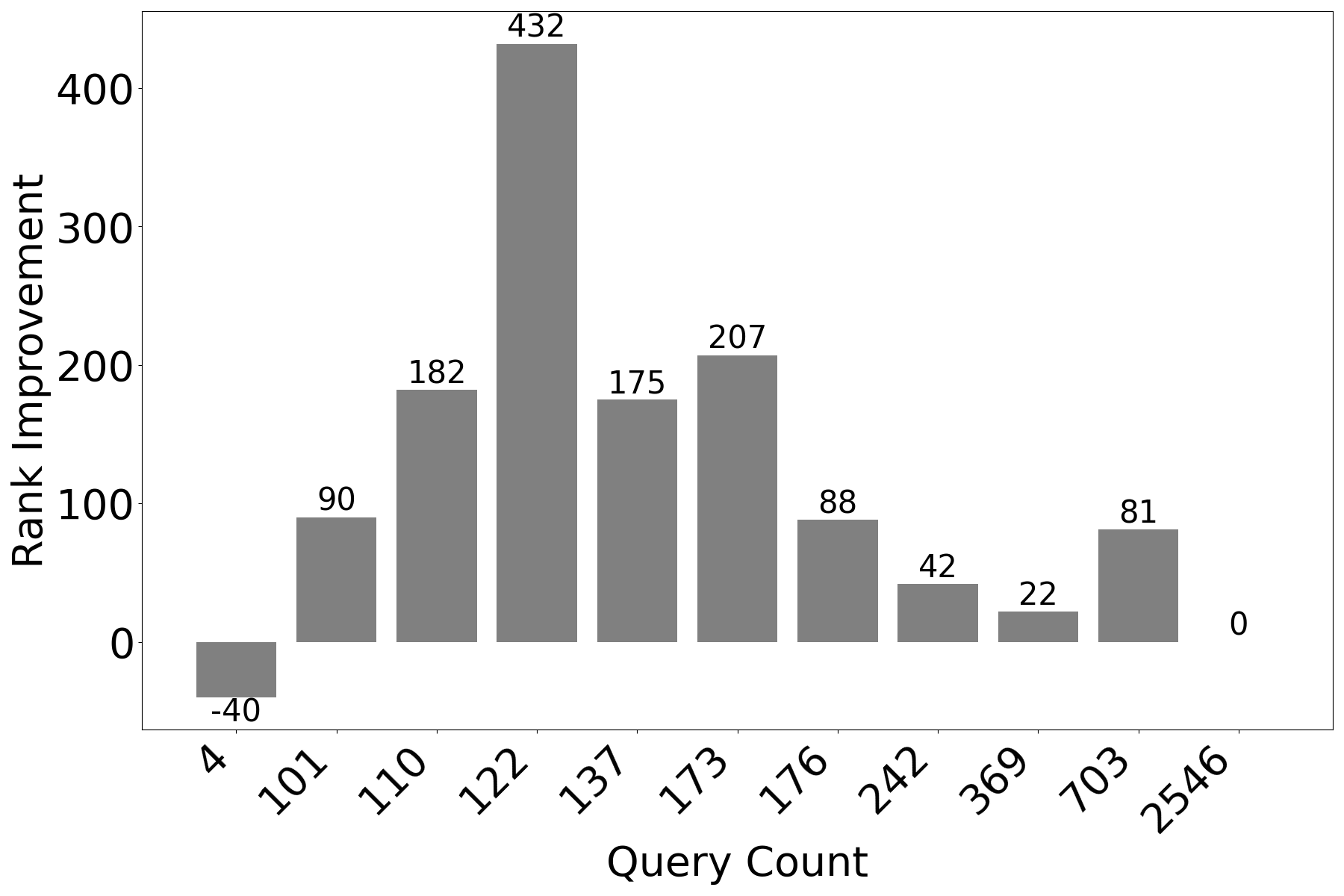}
        \\
        (a) Full Dataset (4,683 Bug Beports)
        \\[0.2cm]
        \includegraphics[width=0.8\linewidth]{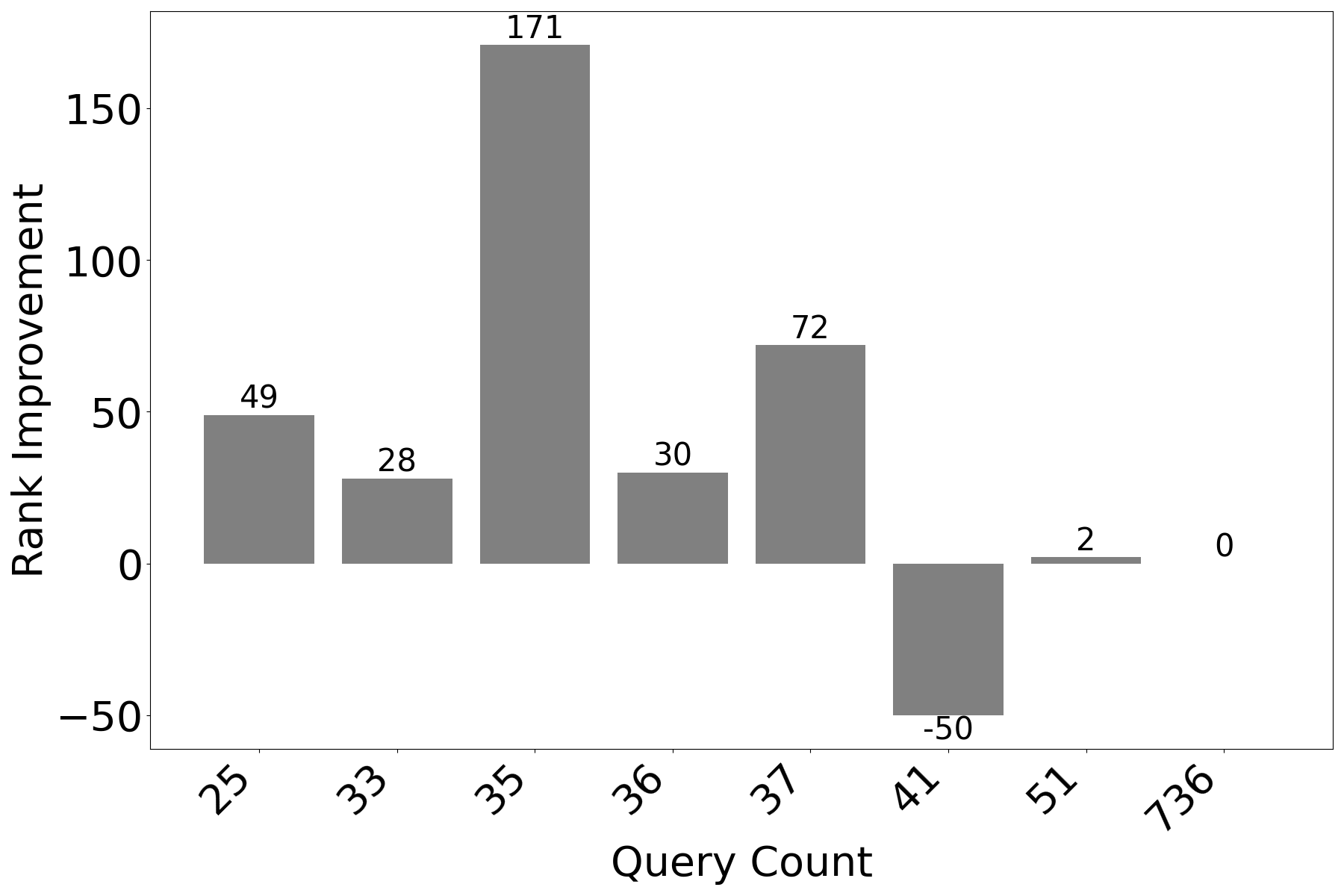}
        \\
        (b) Low Quality Bug Reports (1,101 Bug Reports)

    \end{tabular}
    \caption{Rank Improvement: BRaIn vs Blizzard}
    \label{fig.rankimprovement}
\end{figure}

\begin{table}[ht]
\caption{Statistical Test: BRaIn vs. Blizzard }
\centering
\begin{tabular}{|l|c|c|}
\hline
Evaluation Point & p-value & Effect Size (Cliff’s $\delta$) \\
\hline \hline
Top-1 & 0.0023 ** & Medium (0.41) \\ \hline
Top-5 & 0.0015 ** & Large (0.66) \\ \hline
Top-10 & 0.0008 *** & Large (0.82) \\ \hline
\end{tabular}
\label{tab:statistical_test}
\caption*{*=statistical significance}

\end{table}

Finding the first buggy document is very important during bug localization\cite{prf_haiduc_4}. We further investigate how BRaIn performs in such a case. We chose Blizzard for this investigation as it is the best-performing model against BRaIn in our experiments. Fig. \ref{fig.rankimprovement}-a compares BRaIn and Blizzard by analyzing the difference in ranks for each query within the top 10 results. A positive value indicates that BRaIn found the first ground truth at a better rank than Blizzard, while a negative value suggests the opposite. For 122 bug reports, the large difference of 432 indicates that BRaIn identified the first buggy documents more often than Blizzard. In contrast, Blizzard only outperformed BRaIn for 4 bug reports. For the set of 2,546 bug reports, there are two distinct possibilities: Either (a) the rank difference between the two techniques was 0, or (b) neither technique identified a buggy document within the top-10 results.
We extend our analysis to the 1,101 low-quality bug reports discussed in $RQ_1$. Here we also see BRaIn's dominance in rank improvement over Blizzard for 80.34\% low-quality bug reports (Fig. \ref{fig.rankimprovement}-b).
These results strongly indicate the superiority of our approach. To further validate our findings against Blizzard, we conducted non-parametric statistical tests --Mann-Whitney Wilcoxon and Cliff’s $\delta$\cite{wilcoxon_statistical}-- and compared BRaIn to Blizzard in identifying the first buggy document on the entire dataset. The tests in Table \ref{tab:statistical_test} show $p-values < 0.05$ for the top-1, 5, and 10 results, with medium to large effect sizes (i.e., $0.41 \leq \delta \leq 0.82$). Thus, it confirms BRaIn's consistent ability to detect the first buggy document more effectively than its closest competitor.

\begin{tcolorbox}[colback=lightergray,colframe=black,arc=1mm,boxrule=0.5pt]
\textbf{RQ3 Summary:} BRaIn outperforms traditional, relevance feedback-based, and ML-based baseline techniques, achieving an 87.6\% improvement in MAP. This demonstrates BRaIn's ability to localize relevant buggy documents at the top positions using Intelligent Relevance Feedback (IRF), surpassing relevance feedback-based techniques by 3.7–13.8\% across various metrics. Statistical significance tests further confirm BRaIn's superiority.
\end{tcolorbox}

\section{Related Work}
\subsection{IR based Bug Localization}
Bug localization techniques can broadly be classified into two main groups: spectra-based and information retrieval (IR)-based approaches \cite{ir_bug_localization1}. Spectra-based methods use program execution traces and test methods to localize bugs, making them complex and expensive\cite{spectra1, spectra2}. In contrast, IR-based techniques rely on textual overlap between bug reports and source code to localize the bugs.

Traditional IR-based bug localization methods that leverage the vector space model (VSM) \cite{vector_space_model}, have been enhanced by integrating additional contexts, such as bug report history, code modifications, and version history \cite{saha2014effectiveness_bug_rep_history, wen2016locus, sisman2012incorporating}. For instance, Saha et al. \cite{saha_bleuir} leverage bug report and source code structures, capture eight components from the code and bug reports, and perform eight pairwise searches using a sophisticated retrieval technique, Indri\cite{Indri}. On the other hand, BugLocator \cite{ir_localization_lda_buglocator} combines a modified VSM (rVSM\cite{ir_localization_lda_buglocator}) score with previous bug fix history to improve bug localization. AmaLgam \cite{ir_ml_amalgam} integrates BLUiR, BugLocator, and version history to better detect buggy documents. AmaLgam+ \cite{wang_amalgam+} further incorporates stack traces and bug reporter history, refining bug localization across five ranking components. While advanced, computationally expensive methodologies such as LSI or LDA \cite{ir_localization_lda_buglocator, ir_bug_localization4} are available, their bug localization effectiveness is similar to that of more basic methods\cite{bench4bl}.

In our work, we use a textual similarity-based retrieval with BM25 in Elasticsearch. However, it was complemented by contextual understanding of bugs and Intelligent Relevance Feedback, leveraging the capabilities of LLMs.

\subsection{Query Reformulation}
Poorly constructed queries from software bug reports can significantly hinder IR-based bug localization\cite{mills2020relationship, rahman2021forgotten}. To tackle this issue, researchers have developed query reformulation techniques that can improve search queries by incorporating better terms or eliminating unnecessary ones. For instance, Refoqus\cite{prf_haiduc_4} uses query characteristics and machine learning to recommend strategies like query reduction or expansion for a given query.
Graph-based methods analyze semantic and syntactic relationships within bug reports to identify key terms. Rahman and Roy\cite{blizzard} create text graphs to collect important terms based on three different bug types to improve query reformulation. The authors later demonstrate generating optimal queries in bug reports by employing Genetic algorithms\cite{rahman2021forgotten}, which iteratively refine queries based on search results. Gay et al.\cite{prf_haiduc_4} used Rocchio’s algorithm to improve queries with developer feedback, while Sisman et al.\cite{prf_sisman_kak_1} expanded queries by selecting terms from the top-ranked documents using Spatial Code Proximity (SCP), without using any explicit relevance feedback.

These approaches rely on statistical properties or co-occurrence relations to reformulate an original query without meaningful knowledge of source code or bug reports. Our approach digs deeper to select relevant source code by contextually understanding bug reports to formulate queries for better bug resolution.

\subsection{Deep Learning and Bug Localization}
Recent advancements in deep learning have encouraged its applications in bug localization. DNNLOC\cite{dnnloc_ir}, a seminal work on this topic, identifies buggy documents by learning from multiple text-based features and metadata (e.g., rVSM score\cite{ir_localization_lda_buglocator}, class name similarity, bug report recency). However, its reliance on features like bug fixing recency can limit its application \cite{ir_localization_lda_buglocator}.
A recent technique, FBL-BERT \cite{ciborowska_fbl_bert} uses a BERT-based model, ColBERT \cite{khattab2020colbert}, for document scoring with late interaction. However, it relies on changesets for resolution, which are difficult to track in large, fast-changing projects. Another recent technique, RLocator\cite{RLocator}, optimizes ranking metrics in the bug localization process. It formulates bug localization as a Markov Decision Process (MDP) and employs reinforcement learning (RL) to localize bugs. Other approaches like TRANP-CNN \cite{TRANP-CNN} and CooBa \cite{zhu2020cooba} use Convolutional Neural Networks (CNN) and Graph Convolutional Networks (GCN) to improve cross-project bug localization. However, these methods may struggle with scalability when handling large volumes of documents.

In contrast, our technique ensures scalability by using a limited set of documents retrieved via Elasticsearch\cite{elasticsearch}. By combining efficient IR-based filtering with the contextual depth of language models, BRaIn accurately ranks buggy documents, improving bug localization.

\section{Threats to Validity}
Threats to internal validity concern experimental errors and biases. Replication of existing baselines poses such a threat. We mitigated this by using replication packages from the original authors (Blizzard\cite{BLIZZARD_github}, RLocator\cite{RLocator_replication}) and from Bench4BL\cite{bench4bl} (BLUiR\cite{saha_bleuir}). Due to Indri's obsolescence, we substituted it with Lucene in the BLUiR replication. On the other hand, we replicated DNNLOC\cite{dnnloc_ir}, NextBug\cite{NextBug}, and Sysman-SCP\cite{prf_haiduc_4}) adhering strictly to the original authors' settings and parameters. To minimize bias, we tested on two distinct datasets and found only a negligible difference compared to baseline performances.

Threats to external validity relate to generalizability. While BRaIn was evaluated only on Java code, the underlying models (e.g., LLaMa\cite{llama}) are designed to adapt to various programming languages, potentially mitigating this limitation.

Threats to construct validity concern the appropriateness of our evaluation metrics. We employed widely used metrics such as Mean Average Precision (MAP), Mean Reciprocal Rank (MRR), and HIT@K, which were commonly used in existing literature on bug localization \cite{blizzard, saha_bleuir, Rack} and Information Retrieval studies\cite{evaluation_ir}. Therefore, this choice of metrics minimizes threats to construct validity.

Finally, we used 20 bug reports from our dataset to optimize prompts with LLaMA. Since they are part of our experimental dataset, it could introduce bias. We repeated a limited experiment and found similar performance to the reported ones in the paper. Thus, any relevant threat of bias is minimal.

\section{Conclusion and Future Work}
Software bugs consume resources and cause financial losses\cite{BI_crowdstrike}. Resolving these challenges has been a major focus, with researchers working on solutions for decades. In this paper, we introduced BRaIn to address the contextual gaps between bug reports and source codes by assessing the relevance between bug reports and code with Large Language Models (LLM). Our proposed technique leverages the LLM’s feedback (a.k.a., Intelligent Relevance Feedback) to improve bug localization by reformulating queries and reranking source documents. We evaluated BRaIn’s performance using three widely used metrics in bug localization: Mean Average Precision (MAP), Mean Reciprocal Rank (MRR), and HIT@K. BRaIn achieved substantial improvements, showing increases of 87.6\% in MAP, 89.5\% in MRR, and 48.8\% in HIT@K compared to baseline techniques.

Building on these results, in the future we aim to localize bugs using in few-shot settings\cite{prompt_survey} and at a much finer level (i.e., method, line) to assist developers in pinpointing the exact location of bug inducing code. 

%
%
%
\bibliographystyle{IEEEtran}
\bibliography{IEEEabrv,bibliography}
\balance
\end{document}